\documentclass[preprint,12pt]{elsarticle}

\newtheorem{proposition}{Proposition}

\usepackage{amssymb}
\usepackage{amsmath}

\journal{Physica A}

\begin{document}

\begin{frontmatter}

\title{First and second order semi-Markov chains for wind speed modeling}

\author{Guglielmo D'Amico}
\address{Dipartimento di Farmacia, Universit\`a G. D'Annunzio, 66013 Chieti, Italy}
\author{Filippo Petroni}
\address{Dipartimento di Scienze Economiche ed Aziendali, Universit\`a di Cagliari, 09123 Cagliari, Italy}%

\author{Flavio Prattico}
\address{Dipartimento di Ingegneria Meccanica, Energetica e Gestionale, Universit\`a degli studi dell'Aquila, 67100 L'Aquila, Italy}


\begin{abstract}
The increasing interest in renewable energy, particularly in wind, has given rise to the necessity of accurate models for the generation of good synthetic wind speed data. Markov chains are often used with this purpose but better models are needed to reproduce the statistical properties of wind speed data. 
We downloaded a database, freely available from the web, in which are included wind speed data taken from L.S.I. -Lastem station (Italy) and sampled every 10 minutes. With the aim of reproducing the statistical properties of this data we propose the use of three semi-Markov models. We generate synthetic time series for wind speed by means of Monte Carlo simulations. The time lagged autocorrelation is then used to compare statistical properties of the proposed models with those of real data and also with a synthetic time series generated though a simple Markov chain. 
\end{abstract}

\date{\today}

\end{frontmatter}

\section{Introduction}
\label{Suno}

The increasing interest in renewable energy leads scientific research to find a better way to recover most of the available energy. Particularly, the maximum energy recoverable from wind is equal to 59.3\% of that available (Betz law) at a specific pitch angle and when the ratio between the wind speed in output and in input is equal to $1/3$. The pitch angle is the angle formed between the airfoil of the blade of the wind turbine and the wind direction. Old turbine and a lot of that actually marketed, in fact, have always the same invariant geometry of the airfoil. This causes that wind turbines will work with an efficiency that is lower than 59.3\%. New generation wind turbines, instead, have a system to variate the pitch angle by rotating the blades. This system allows the wind turbines to recover, at different wind speed, always the maximum energy, working in Betz limit at different speed ratios. A powerful system control of the pitch angle allows the wind turbine to recover better the energy in transient regime. A good stochastic model for wind speed is then needed to help both the optimization of turbine design and to assist the system control to predict the value of the wind speed to positioning the blades quickly and correctly. The possibility to have synthetic data of wind speed is a powerful instrument to assist designer to verify the structures of the wind turbines or to estimate the energy recoverable from a specific site. To generate synthetic data, Markov chains of first or higher order were often used \cite{sha05,nfa04,you03,kan04}. In particular in \cite{sha05} is presented a comparison between a first-order Markov chain and a second-order Markov chain. A similar work, but only for the first-order Markov chain, is conduced by \cite{nfa04}, presenting the probability transition matrix and comparing the energy spectral density and autocorrelation of real and synthetic wind speed data. A tentative to modeling and to join speed and direction of wind is presented in \cite{you03}, by using two models, first-order Markov chain with different number of states, and Weibull distribution.
The Markov assumption is widely used in other closely related topics, see for example \cite{coll05,ivan02}.

All these models use Markov chains to generate synthetic wind speed time series but the search for a better model is still open. Approaching this issue, we applied new models which are generalization of Markov models. More precisely we applied semi-Markov models to generate synthetic wind speed time series.

Semi-Markov processes (SMP) are a wide class of stochastic processes which generalize at the same time both Markov chains and renewal processes \cite{cinl69,limn01}. They have been widely used in the literature to model natural phenomena (see for example \cite{barb04,jans07,dami09,dami09b,dami11,dami12a,dami12b}). Their main advantage is that of using whatever type of waiting time distribution for modeling the time to have a transition from one state to another one. This major flexibility has a price to pay: availability of data to estimate the parameters of the model which are more numerous. Data availability is not an issue in wind speed studies, therefore, semi-Markov models can be used in a statistical efficient way.   

In this work we present three different semi-Markov chain models: the first one is a first-order SMP where the transition probabilities from two speed states (at time $T_n$ and $T_{n-1}$) depend on the initial state (the state at $T_{n-1}$), final state (the state at $T_{n}$) and on the waiting time (given by $t=T_{n}-T_{n-1}$). The second model is a second order SMP where we consider the transition probabilities as depending also on the state the wind speed was before the initial state (which is the state at $T_{n-2}$) and the last one is still a second order SMP where the transition probabilities depends on the three states at $T_{n-2},T_{n-1}$ and $T_{n}$ and on the waiting times $t_1=T_{n-1}-T_{n-2}$ and $t_2=T_{n}-T_{n-1}$. The three models will be extensively explained in the next sections. 
The three models are used to generate synthetic time series for wind speed by means of Monte Carlo simulations and the time lagged autocorrelation function is used to compare statistical properties of the proposed models with those of real data and also with a time series generated though a simple Markov chain. The probability density function of real and simulated data are also compared for the model which is recognized to be the best among the proposed ones. 

In this paper, for the first time, different second order discrete time semi-Markov chains are defined and general formulae of transition probabilities with initial and final backward are presented. 

The paper is organized as follows: in Section 2 we describe the stochastic models and we derive relevant results. Section 3 demonstrates the models applied to a real dataset by testing the semi-Markov hypothesis and by computing the autocorrelation functions and the probability density functions of the wind speed. Finally Section 4 presents some concluding remarks.

\section{Wind speed modeling with semi-Markov chains}

Semi-Markov chains are a generalization of Markov chains allowing the times between transitions to occur at random times according to any kind of distribution functions which may depend on the current and the next visited state. As it is well known, Markov chains have sojourn times between transitions geometrically distributed, for this reason the memoryless property is preserved and no duration effect is observed. The more general semi-Markov environment allows the possibility to use also no memoryless distributions and then can reproduce a duration effect. The duration effect affirms that the time the system is in a state influences its transition probabilities. The states of the process in our data are represented by different wind speed values, then in this paper we detect the presence of a duration effect in wind speed modeling and forecasting.
Given that the data on wind speed are recorded naturally in discrete time and in discrete values we develop also the theoretical model in discrete time and discrete values.\\
\indent Here below we propose a semi-Markov model of order two in state and duration and we compare its performance with the Markov chain models often used to describe wind speed, see \cite{sha05,nfa04,you03} and with some particular cases of our semi-Markov chain model.\\
Let us consider a finite set of states $E=\{1,2,...,S\}$ in which the system can be into and a complete probability space $(\Omega, \emph{F}, P)$ on which we define the following random variables:
\begin{equation}
\label{uno}
J_{n}:\Omega\rightarrow E, \,\,\, T_{n}:\Omega\rightarrow \mathbb{N} .
\end{equation}
\indent They denote the state occupied at the n-th transition and the time of the n-th transition, respectively. To be more concrete, by $J_{n}$ we denote the wind speed at the nth transition and by $T_{n}$ the time of the nth transition of the wind speed process. We do the following conditional independence assumption:
\begin{equation}
\label{ventuno}
\begin{aligned}
& P[J_{n+1}=j,T_{n+1}-T_{n}= t |\sigma(J_{s},T_{s}), J_{n}=k, J_{n-1}=i, T_{n}-T_{n-1}=x, 0\leq s \leq n]\\
& \quad =P[J_{n+1}=j,T_{n+1}-T_{n}= t |J_{n}=k, J_{n-1}=i,T_{n}-T_{n-1}=x ]:=\,\,_{x}q_{i.k,j}(t).
\end{aligned}
\end{equation}
\indent Relation $(\ref{ventuno})$ asserts that, the knowledge of the values $J_{n}, J_{n-1},T_{n}-T_{n-1}$ is sufficient to give the conditional distribution of the couple $J_{n+1}, T_{n+1}-T_{n}$ whatever the values of the past variables might be. Therefore to make probabilistic forecasting we need the knowledge of the last two visited state and the duration time of the transition between them. For this reason we called this model a second order semi-Markov chains in state and duration.\\
\indent The conditional probabilities 
$$
_{x}q_{i.k,j}(t)=P[J_{n+1}=j, T_{n+1}-T_{n}=t|J_{n}=k,J_{n-1}=i,T_{n}-T_{n-1}=x ]
$$ 
are stored in a matrix of functions $\mathbf{q}=(_{x}q_{i.k,j}(t))$ called the second order kernel in state and duration. The element $_{x}q_{i.k,j}(t)$ represents the probability that next wind speed will be in speed $j$ at time $t$ given that the current wind speed is $k$ and the previous wind speed state was $i$ and the duration in wind speed $i$ before of reaching the wind speed $k$ was equal to $x$ units of time.\\  
\indent We can define the cumulated second order kernel:
\begin{equation}
\label{ventidue}
\begin{aligned}
& _{x}Q_{i.k,j}(t):=P[J_{n+1}=j, T_{n+1}-T_{n}\leq t|J_{n}=k,J_{n-1}=i,T_{n}-T_{n-1}=x]\\
& =\sum_{s=1}^{t} {_{x}q_{i.k,j}(s)}.
\end{aligned}
\end{equation}
\indent The process $\{J_{n}\}$ is a second order Markov chain with state space $E$ and transition probability matrix $_{x}\mathbf{P}=\,_{x}\mathbf{Q}(\infty)$. We shall refer to it as the embedded Markov chain.\\
\indent Let us consider the unconditional waiting time distribution function in states $i$ and $k$ with duration $x$ as
\begin{equation}
\label{ventitre}
_{x}H_{i.k}(t):=P[T_{n+1}-T_{n}\leq t|J_{n}=k,J_{n-1}=i,T_{n}-T_{n-1}=x]=\sum_{j\in E} {_{x}Q_{i.k,j}(t)}.
\end{equation}
\indent The conditional cumulative distribution functions of the waiting time in each state, given the state subsequently occupied is defined as
\begin{equation}
\label{ventiquattro}
\begin{aligned}
& _{x}G_{i.k,j}(t)=P[T_{n+1}-T_{n}\leq t|J_{n}=k,J_{n-1}=i, J_{n+1}=j,T_{n}-T_{n-1}=x]\\
& =\frac{1}{_{x}p_{i.k,j}}\sum_{s=1}^{t} {_{x}q_{i.k,j}(s)}\cdot 1_{\{_{x}p_{i.k,j}\neq 0\}}+1_{\{_{x}p_{i.k,j}=0\}},
\end{aligned}
\end{equation}
\noindent where, in general, $1_{\{a=b\}}$ is $1$ if $a=b$ and zero otherwise.\\
\indent Now let $N(t)=\sup\{n:T_{n}\leq t\}$ $\forall t\in \mathbb{N}$ be the number of transitions up to time $t$, then the second order (in state and duration) semi-Markov chain can be defined as $Z(t)=(Z^{1}(t),Z^{2}(t))=(J_{N(t)-1},J_{N(t)}))$.\\
\indent For all $i,k,j\in E$, and $t\in \mathbb{N}$, we define the semi-Markov transition probabilities:
\begin{equation}
\label{venticinque}
_{x}\phi_{i.k,h.j}(t):=P[J_{N(t)}=j,J_{N(t)-1}=h|J_{N(0)}=k,J_{N(0)-1}=i,T_{N(0)}=0,T_{N(0)}-T_{N(0)-1}=x],
\end{equation}
\noindent then the following system of equations is verified:
\begin{equation}
\label{ventisei}
_{x}\phi_{i.k,h.j}(t)=1_{\{i=h,k=j\}}(1-\,  _{x}H_{i.k}(t))+\sum_{r\in
E}\sum_{s =1}^{t}\, _{x}q_{i.k,r}(s)\, _{s}\phi_{k.r,h.j}(t-s).
\end{equation}
\indent The proof of equation $(\ref{ventisei})$ is not given here because it is a particular case of the equation established and proved later.\\ 
\indent To detect the duration effects let us introduce the backward recurrence time process defined for each time $t\in \mathbb{N}$ by:
\begin{equation}
\label{otto}
   B(t)=t-T_{N(t)}.
\end{equation}
\indent If the semi-Markov process $Z(t)$ indicates the wind speed at time $t$, the backward process $B(t)$ indicates the time since the last transition, that is from how long time the wind speed is at the value $Z(t)$.

\begin{figure}
\centering
\includegraphics[height=4cm]{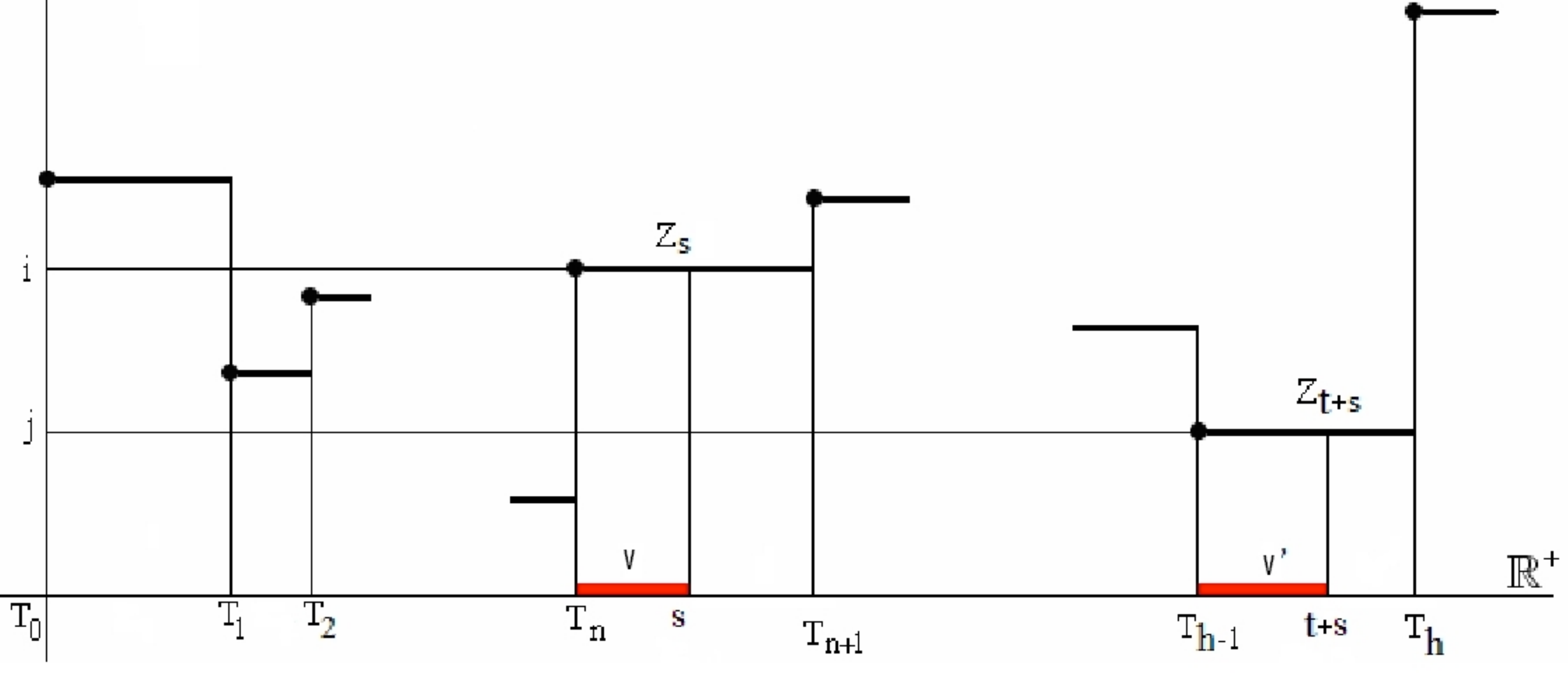}
\caption{Initial and final backward values}\label{fig0}
\end{figure}

\indent In Figure \ref{fig0} a trajectory of a semi-Markov process with initial and final backward times is reported. In this figure we have that $N(s)=n$ and $N(s+t)=h-1$, the backward process at time $s$ is $B(s)=s-T_{n}=v$ and the backward process at time $t+s$ is $B(t+s)=t+s-T_{h-1}=v'$.\\
\hspace{-1cm}
\indent To quantify the duration effect in our second order semi-Markov model, let us define the following probabilities:
\begin{equation}
\label{ventisette}
\begin{aligned}
& _{x}^{b}\phi_{i.k,h.j}^{b}(v;v',t):=\\
& P[J_{N(t)}=j,B(t)=v',J_{N(t)-1}=h|J_{N(0)}=k,J_{N(0)-1}=i,B(0)=v,T_{N(0)}-T_{N(0)-1}=x].
\end{aligned}
\end{equation}
\indent Expression $(\ref{ventisette})$ gives the probability that the wind speed will enter in the state $j$ at time $t-v'$ coming from state $h$ and will remains inside the state $j$ without any other transition up to the time $t$ given that at the present the wind speed is $k$ and it entered into this state with the last transition $v$ periods before coming from a wind speed equal to $i$ with a duration in $i$ of $x$ periods.
\begin{proposition}
The relation $(\ref{ventotto})$ represents the evolution equation of $(\ref{ventisette})$:
\begin{equation}
\label{ventotto}
\begin{aligned}
& _{x}^{b}\phi_{i.k,h,j}^{b}(v;v',t)=1_{\{i=h,k=j,v'=t+v\}}\frac{[1-\, _{x}H_{i.k}(t+v)]}{[1-\, _{x}H_{i.k}(v)]}\\
& +\sum_{r\in E}\sum_{s=1}^{t-v'}\frac{_{x}q_{i.k,r}(s+v)}{[1-\, _{x}H_{i.k}(v)]}\,\,_{s+v}^{\,\,\,\,\,\,\,\,b}\phi_{k.r,h.j}^{b}(0;v',t-s).
\end{aligned}
\end{equation}
\end{proposition}
{\bf{Proof}}: For simplicity of notation, suppose the counting process $N(0)=0$, and denote by
\[
_{x}A_{i.k}(v;0)=\{J_{0}=k,J_{-1}=i,B(0)=v,T_{0}-T_{-1}=x\}.
\]
\indent Then we have
\begin{equation}
\label{primo}
\begin{aligned}
& _{x}^{b}\phi_{i.k,h,j}^{b}(v;v',t)\\
& =P[J_{N(t)}=j,B(t)=v',J_{N(t)-1}=h, T_{1}>t|\,_{x}A_{i.k}(v;0)]\\
& +P[J_{N(t)}=j,B(t)=v',J_{N(t)-1}=h,T_{1}\leq t|\,_{x}A_{i.k}(v;0)].
\end{aligned}
\end{equation}
\indent Observe that 
\begin{equation}
\begin{aligned}
& P[J_{N(t)}=j,B(t)=v',J_{N(t)-1}=h, T_{1}>t|\,_{x}A_{i.k}(v;0)]\\
& =P[J_{N(t)}=j,B(t)=v',J_{N(t)-1}=h|T_{1}> t,\,_{x}A_{i.k}(v;0)]\\
& \cdot P[T_{1}> t|\,_{x}A_{i.k}(v;0)].
\end{aligned}
\end{equation}
If $T_{1}>t$ then $J_{N(t)}=J_{0}$, $J_{N(t)-1}=J_{-1}$ and $B(t)=v'=v+t$. This gives:
\begin{equation}
\label{secondo}
\begin{aligned}
& P[J_{N(t)}=j,B(t)=v',J_{N(t)-1}=h, T_{1}>t|\,_{x}A_{i.k}(v;0)]\\
& =P[k=j, v'=v+t, i=h|T_{1}> t,\,_{x}A_{i.k}(v;0)]\\
& \cdot P[T_{1}> t|J_{0}=k,J_{-1}=i,T_{0}-T_{-1}=x,T_{0}=-v,T_{1}>0]\\
& = 1_{\{k=j,i=h,v'=v+t\}}\frac{P[T_{1}> t|J_{0}=k,J_{-1}=i,T_{0}-T_{-1}=x,T_{0}=-v]}{P[T_{1}>0|J_{0}=k,J_{-1}=i,T_{0}-T_{-1}=x,T_{0}=-v]}\\
& = 1_{\{k=j,i=h,v'=v+t\}}\frac{1- _{x}H_{i.k}(t+v)}{1- _{x}H_{i.k}(v)}.
\end{aligned}
\end{equation}
The second addend on the right hand side of $(\ref{primo})$ can be represented as follows:
\begin{equation}
\label{terzo}
\begin{aligned}
& \sum_{r\in E}\sum_{s=1}^{t-v'}P[J_{N(t)}=j,B(t)=v',J_{N(t)-1}=h,J_{1}=r,T_{1}=s|\,_{x}A_{i.k}(v;0)]\\
& =\sum_{r\in E}\sum_{s=1}^{t-v'}P[J_{N(t)}=j,B(t)=v',J_{N(t)-1}=h|J_{1}=r,T_{1}=s,\,_{x}A_{i.k}(v;0)]\\
& \cdot P[J_{1}=r,T_{1}=s|\,_{x}A_{i.k}(v;0)]\\
& = \sum_{r\in E}\sum_{s=1}^{t-v'}P[J_{N(t)}=j,B(t)=v',J_{N(t)-1}=h|\,_{s+v}A_{k.r}(0;s)]\\
& \cdot P[J_{1}=r,T_{1}-T_{0}=s+v|J_{0}=k,J_{-1}=i,T_{0}=-v, T_{1}>0,T_{0}-T_{-1}=x]\\
& = \sum_{r\in E}\sum_{s=1}^{t-v'} {_{s+v}^{\,\,\,\,\,\,\,\,b}\phi_{k.r,h.j}^{b}(0;v',t-s)}\\
& \cdot \frac{P[J_{1}=r,T_{1}-T_{0}=s+v|J_{0}=k,J_{-1}=i,T_{0}-T_{-1}=x]}{P[T_{1}-T_{0}>v|J_{0}=k,J_{-1}=i,T_{0}-T_{-1}=x]}\\
& =\sum_{r\in E}\sum_{s=1}^{t-v'}\frac{_{x}q_{i.k,r}(s+v)}{[1-\, _{x}H_{i.k}(v)]}\,\,_{s+v}^{\,\,\,\,\,\,\,\,b}\phi_{k.r,h.j}^{b}(0;v',t-s).
\end{aligned}
\end{equation}
\indent A substitution of $(\ref{secondo})$ and $(\ref{terzo})$ in $(\ref{primo})$ concludes the proof.\\

\indent Obviously we have that
\begin{equation}
\label{ventinove}
\begin{aligned}
&  _{x}^{b}\phi_{i.k,h.j}(v;t)\\
& :=P[J_{N(t)}=j,J_{N(t)-1}=h|J_{0}=k,J_{-1}=i,B(0)=v,T_{0}=0,T_{0}-T_{-1}=x]\\
& =\sum_{v'\geq 0} {_{x}^{b}\phi_{i.k,h.j}^{b}(v;v',t)}.
\end{aligned}
\end{equation}
\indent Expression $(\ref{ventinove})$ represents the probability that the wind speed will be in the state $j$ at time $t$ coming from a wind speed equal to $h$ given that at present the wind speed is $k$ and it entered into this state with the last transition $v$ periods before coming from a wind speed equal to $i$ with a duration in $i$ of $x$ periods.\\
\indent Notice that, if $v=0$ we obtain the equation $(\ref{ventisei})$.\\
\indent It should be noted that our semi-Markov model of order two in state and duration contains several interesting special cases we will apply in the next section. The paper \cite{limn03} proposed a n-order semi-Markov process (in state) in continuous time. The discrete time counterpart of order two (in state) is obtained through the following assumption:
\begin{equation}
\label{dodici}
\begin{aligned}
& P[J_{n+1}=j,T_{n+1}-T_{n}= t |\sigma(J_{s},T_{s}), J_{n}=k, J_{n-1}=i, 0\leq s \leq n]\\
& \quad =P[J_{n+1}=j,T_{n+1}-T_{n}= t |J_{n}=k, J_{n-1}=i]:=q_{i.k,j}(t).
\end{aligned}
\end{equation}
\indent Relation $(\ref{dodici})$ asserts that, the knowledge of the values $J_{n}, J_{n-1}$ is sufficient to give the conditional distribution of the couple $J_{n+1}, T_{n+1}-T_{n}$ whatever the values of the past variables might be. Therefore to make probabilistic forecasting we need the knowledge of the last two visited state. In the application we will refer to this model as the model named semi-Markov II.\\
\indent If we assume that
\begin{equation}
\label{due}
\begin{aligned}
& P[J_{n+1}=j,T_{n+1}-T_{n}= t |\sigma(J_{s},T_{s}), J_{n}=i, 0\leq s \leq n]\\
& \quad =P[J_{n+1}=j,T_{n+1}-T_{n}= t |J_{n}=i]:=q_{ij}(t).
\end{aligned}
\end{equation}
then we recover the classical semi-Markov chain model. Finally remark that a Markov chain with transition probability matrix $(t_{ij})_{ij\in E}$ is obtained when the semi-Markov kernel $(\ref{due})$ is given by $q_{ij}(s)=t_{ij}(t_{ii})^{s-1}1_{\{i\neq j\}}$, $s\in \mathbb{N} -\{0\}$.

\section{Application to real data}
To check the validity of our model we perform a comparison of the behavior of real data
and wind speeds generated through Monte Carlo simulations based on the models described above.
In this section we describe the database of real data used for the analysis, the method used to simulate
synthetic wind speed time series and, at the end, we compare results from real and simulated data.   

\begin{figure}
\centering
\includegraphics[height=8cm]{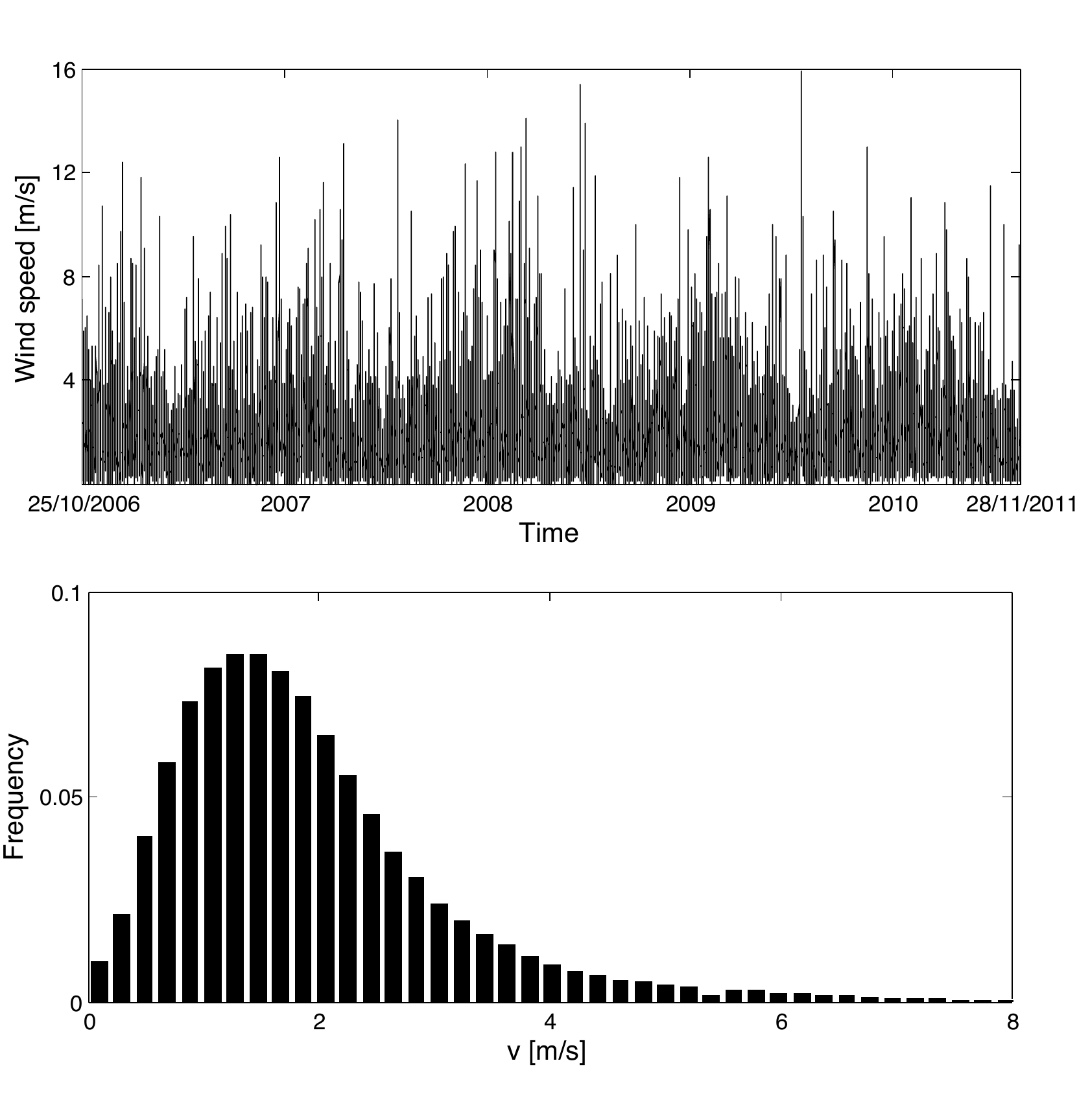}
\caption{Time series of wind speed and its empirical distribution.}\label{fig1}
\end{figure}

The data used in this analysis are freely available from $http://www.lsi-lastem.it/meteo/page/dwnldata.aspx$. The station of L.S.I. -Lastem is situated in Italy at N 45$°$ 28' 14,9'' $-$ E 9$°$ 22' 19,9'' and at 107 $m$ of altitude. The station uses a combined speed-direction anemometer at 22 $m$ above the ground. It has a measurement range that goes from 0 to 60 $m/s$, a threshold of 0,38 $m/s$ and a resolution of 0,05 $m/s$. The station processes the speed every 10 minutes in a time interval ranging from 25/10/2006 to 28/06/2011. 
During the 10 minutes are performed 31 sampling which are then averaged in the time interval.
In this work, we use the sampled data that represents the average of the modulus of the wind speed ($m/s$) without a specific direction.
The database is then composed of about 230thousands wind speed measures ranging from 0 to 16 $m/s$. The time series, together with its empirical probability density distribution are represented in Figure \ref{fig1}.

To be able to model the wind speed as a semi-Markov process the state space of wind speed has been discretized.
In the example shown in this work we discretized wind speed into 7 states chosen to cover all the wind speed distribution. 
From the discretized wind speeds we estimated the probabilities ${\bf P}$ and $G$ to generate synthetic trajectories by means of Monte Carlo simulations \cite{gis11} for three semi-Markov models: a simple semi-Markov model of the first order named semi-Markov I, semi-Markov II a second order semi-Markov model in state (as described in section 2) and the second order semi-Markov model in state and duration is named semi-Markov III.  For comparison reason, we also generated a synthetic trajectory which follows a simple Markov model with transition probability matrix estimated from the real data. We then ended up with five trajectories: one representing real data, three representing the three trajectories according to the three semi-Markov models and the last one a Markov chain.
The time series are used in the following to compare results on the time lagged autocorrelation function. Real data are, in fact, strongly autocorrelated and the autocorrelation function decreases rapidly with time. We then tested our models to check whether they are able to reproduce such behavior. 
Before doing so we tested the semi-Markov hypothesis using an hypothesis test which we are going to describe in the next subsection.

\subsection{Test}
The semi-Markov hypothesis is tested applying a test of hypothesis proposed by \cite{ste06} and shortly described here below. As already stated, the model can be considered semi-Markovian if the sojourn times are not geometrically distributed. The probability distribution function of the sojourn time in state $i$ before making a transition in state $j$ has been denoted by $G_{ij}(\cdot)$. Define the corresponding probability mass function by 
\begin{eqnarray}
\label{pmf}
&&g_{ij}(t)=P\{T_{n+1}-T_{n}= t|J_{n}=i, J_{n+1}=j\}=\nonumber \\ 
&&\left\{
                \begin{array}{cl}
                       \ G_{ij}(t)-G_{ij}(t-1)  &\mbox{if $t > 1$}\\
                         G_{ij}(1)  &\mbox{if $t=1$}\\
                   \end{array}
             \right.
\end{eqnarray}
Under the geometrical hypothesis the equality $g_{ij}(1)(1-g_{ij}(1))-g_{ij}(2)=0$ must hold, then a sufficiently strong deviation from this equality has to be interpreted as an evidence in favor of the semi-Markov model. The test-statistic is the following:
\begin{equation}
\label{test}
\hat{S}_{ij}=\frac{\sqrt{N(i,j)}\big(\hat{g}_{ij}(1)(1-\hat{g}_{ij}(1))-\hat{g}_{ij}(2)\big)}{\sqrt{\hat{g}_{ij}(1)(1-\hat{g}_{ij}(1))^{2}(2-\hat{g}_{ij}(1))}}.
\end{equation}
\noindent where $N(i,j)$ denotes the number of transitions from state $i$ to state $j$ observed in the sample and $\hat{g}_{ij}(x)$ is the empirical estimator of the probability $g_{ij}(x)$ which is given by the ratio between the number of transition from $i$ to $j$ occurring exactly after $x$ unit of time and $N(i,j)$. This statistic, under the geometrical hypothesis $H_{0}$ (or markovian hypothesis), has approximately the standard normal distribution, see \cite{ste06}.\\
\indent We applied this procedure to our data to execute tests at a significance level of $95\%$. Because we have $7$ states we estimated the  $7\times (7-1)$ waiting time distribution functions and for each of them we computed the value of the test-statistic (\ref{test}). The geometric hypothesis is rejected for $28$ of the $42$ distributions. Due to lack of space, we do not report all the values of the test-statistic, but they
are available upon request. In Table 1 we show the results of the test applied to the waiting time distribution functions for few states.\\
\begin{table}
\label{table1}  
\begin{center}
\begin{tabular}{llll}
\hline\noalign{\smallskip}
\textbf{state} & \textbf{state} & \textbf{score} & \textbf{decision} \\
\noalign{\smallskip}\hline\noalign{\smallskip}
$i=1$ & $j=2$ & 12.14 & $H_{0}$ rejected \\
$i=1$ & $j=3$ & 3.33 & $H_{0}$ rejected \\
$i=2$ & $j=1$ & 5.64 & $H_{0}$ rejected \\
$i=2$ & $j=3$ & 9.35 & $H_{0}$ rejected \\
\noalign{\smallskip}\hline
\end{tabular}
\caption{Results of the Test}
\end{center}
\end{table}
\indent The large values of the test statistic suggest the rejection of the Markovian hypothesis in favor of the more general semi-Markov one.\\

\subsection{Autocorrelation function}
If $Z$ indicates wind speed, the time lagged $(\tau)$ autocorrelation of wind speed is defined as 
\begin{equation}
\label{autosquare}
\Sigma(\tau)=\frac{Cov(Z(t+\tau),Z(t))}{Var(Z(t))}
\end{equation}
The time lag $\tau$ was made to run from 1 minute up to 1000 minutes. Note that to be able to compare results for $\Sigma(\tau)$ each
simulated time series was generated with the same length as real data.
\begin{figure}
\centering
\includegraphics[height=8cm]{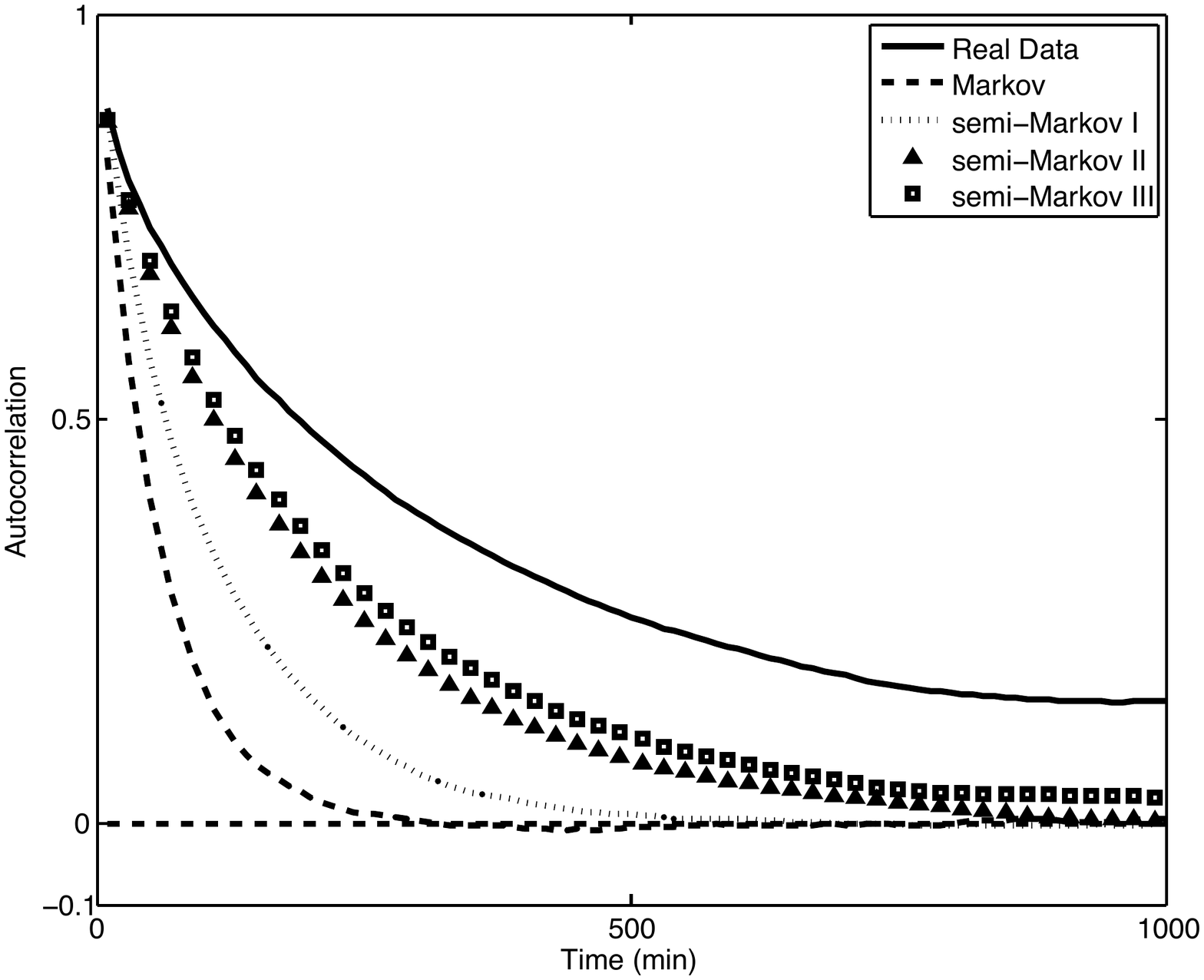}
\caption{Time lagged autocorrelation function.}\label{fig2}
\end{figure}
Results shown in Figure \ref{fig2} indicate that semi-Markov models reproduce better the autocorrelation present in real data especially if a second order semi-Markov chain is used to generate synthetic data. We can also notice that the second order semi-Markov model in state and duration (Model III) has to be preferred to the second order semi-Markov model in state (Model II) because it exhibits a slightly higher autocorrelation uniformly in time and also because, asymptotically,  its autocorrelation has the same slope of that observed in real data whereas in the Model II, the autocorrelation drops to zero. 

\subsection{Probability density function and trajectories}
\begin{figure}
\centering
\includegraphics[height=8cm]{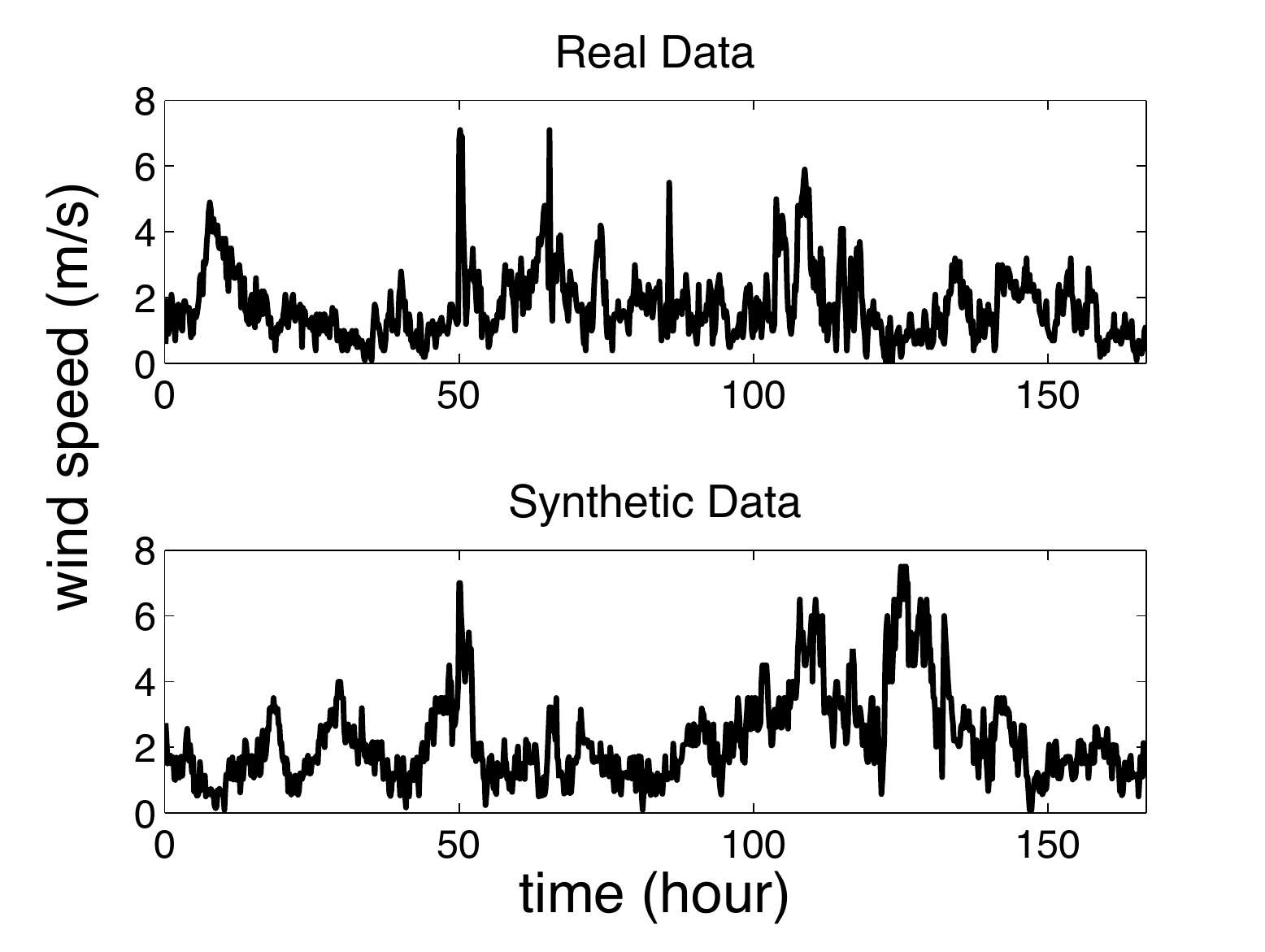}
\caption{Trajectories for 160 hours for real and synthetic data. Data are sampled every 10 minutes in both cases.}\label{fig3}
\end{figure}

\begin{figure}
\centering
\includegraphics[height=8cm]{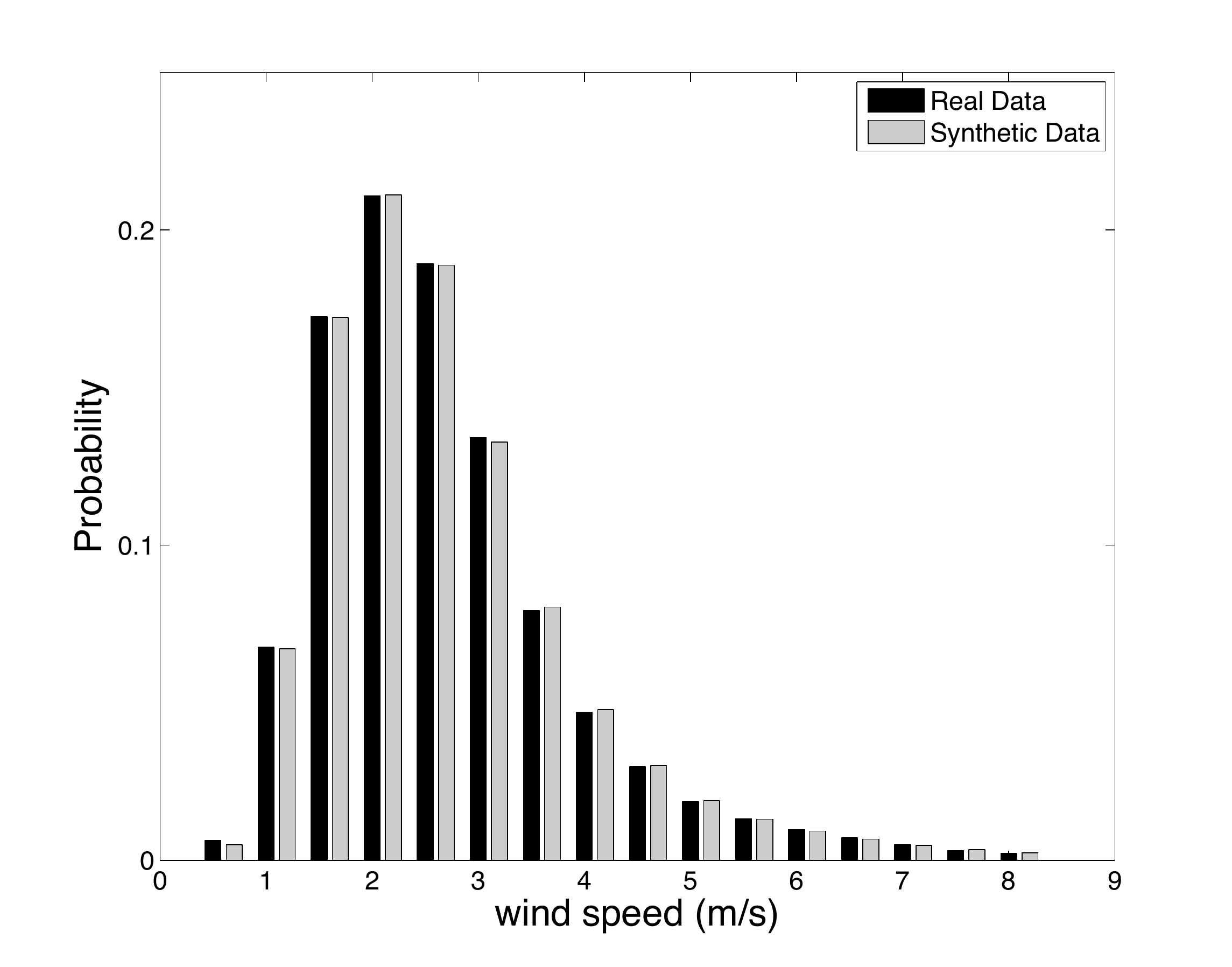}
\caption{Probability density function for real and synthetic data}\label{fig4}
\end{figure}

From the study of the autocorrelation function we found out that Model III should be chosen to generate synthetic trajectories of wind speed data. In this subsection we are going to compare the probability density function (pdf) of real data and synthetic data. As an example we will show also the two trajectories for a sample period.
First of all let us remind that to use the semi-Markov approach we discretized the wind speed in a finite number of states. We choose $7$ so that we can cover all the wind speeds in our dataset and, at the same time, each state as a sufficient number of occupation. To obtain synthetic wind speeds data comparable with real ones a transformation back to continuous speed as to be made. We used the following formula
\begin{equation}
w_{c}(t)=w_{d}(t)+\epsilon \Delta.
\end{equation}
where $w_{c}$ indicates the continuous wind speed at time $t$, $w_{d}(t)$ the discretized wind speed at the same time, $\epsilon$ is a uniformly distributed random number in the interval $[0,1]$ and $\Delta$ is the wind speed interval used for discretization.
We show in Figure \ref{fig3} an example of real and simulated trajectories while in Figure \ref{fig4} we compare the pdf for both trajectories.

From Figure \ref{fig4} it is possible to notice that the pdf for real and synthetic data are almost identical supporting even more our model. We only compare here the pdf of real data and of the synthetic data from Model III because the fitting is already very good and there is very little space for improvement.

\section{Conclusions}
Wind speed is a stochastic process for which a completely satisfactory model is still lacking. 
Many authors have used Markov chain to model the process but this approach does not 
give the same persistence present in real data. We presented in this work three semi-Markov 
models with the aim of generate synthetic wind speed data. We have shown that all our 
models perform better than a simple Markov chain in reproducing the statistical properties 
of wind speed data. In particular,  the model that we recognized as being the more suitable 
is a second order semi-Markov process in state and duration.
Although the evidence shows the semi-Markovian nature of the studied phenomenon, probably a third/fourth order semi-Markov chain would be needed to decrease the difference between autocorrelation of real and simulated data. In our view this approach would be too much computationally and data consuming and further research on a simplified, but still with longer memory, model is needed. In this view, models from the econophysics literature (see for example \cite{dami11,dami12b,pas99a,pas99b,ser06,bav01}) could be tested for possible application in wind speed modeling.

We conclude that semi-Markov models should be used when dealing with wind speed data.

\end{document}